\documentclass[aps,pre,twocolumn,notitlepage,nofootinbib,superscriptaddress]{revtex4-1}
\usepackage{amsmath,amsfonts,amssymb,bm,cancel}
\usepackage{graphicx,float}
\usepackage{hyperref}
\usepackage{enumitem}
\usepackage{soul}
\usepackage[dvipsnames]{xcolor}

\begin{document}

\title{Synchronization of active rotators interacting with environment
}

\author{Taegeun Song}
\affiliation{Department of Physics, Pohang University of Science and Technology (POSTECH), Pohang, 37673, Korea}
\author{Heetae Kim}
\affiliation{Department of Industrial Engineering, Faculty of Engineering, Universidad de Talca, Curic\'o, 3341717, Chile}
\affiliation{Asia Pacific Center for Theoretical Physics (APCTP), Pohang, 37673, Korea}
\author{Seung-Woo Son}
\email{sonswoo@hanyang.ac.kr}
\affiliation{Asia Pacific Center for Theoretical Physics (APCTP), Pohang, 37673, Korea}
\affiliation{Department of Applied Physics, Hanyang University, Ansan, 15588, Korea}
\author{Junghyo Jo}
\email{jojunghyo@snu.ac.kr}
\affiliation{Department of Statistics, Keimyung University, Daegu, 42601, Korea}
\affiliation{School of Computational Sciences, Korea Institute for Advanced Study, Seoul, 02455, Korea}
\affiliation{Department of Physics Education, Seoul National University, Seoul, 08826, Korea}



\begin{abstract}
Multiple organs in a living system respond to environmental changes, and the signals from the organs regulate the physiological environment. Inspired by this biological feedback, we propose a simple autonomous system of active rotators to explain how multiple units are synchronized under a fluctuating environment
.
We find that the feedback via an environment can entrain rotators to have synchronous phases for specific conditions. 
This mechanism is markedly different from the simple entrainment by a common oscillatory external stimulus that is not interacting with systems.
We theoretically examine how the phase synchronization depends on the interaction strength between rotators and environment.
Furthermore, we successfully demonstrate the proposed model by realizing an analog electric circuit with microelectronic devices.
This bio-inspired platform can be used as a sensor for monitoring varying environments, and as a controller for amplifying signals by their feedback-induced synchronization.
\end{abstract}

\maketitle

\section{Introduction}
Living systems maintain their physiological equilibrium for survival, called homeostasis~\cite{homeo1, homeo2}. It literally means `staying the same', and is also an important concept for controllers such as thermostats~\cite{thermostat} in engineering.
Under fluctuating environment with uncertainty, it is crucial to keep dynamical equilibrium for the proper functioning of living systems. 

The regulation of blood glucose levels is one of the most primitive examples of the homeostasis to keep energy balance for living systems.
The islets of Langerhans in the pancreas respond to varying glucose levels, and produce hormones in an oscillatory manner to regulate the glucose homeostasis~\cite{Langerhans}. The phase of hormone oscillation is modulated by the glucose stimulus depending on glucose levels.
As the glucose level increases, the ratio of active to silent phases of the oscillation increases, while its period changes minimally~\cite{hormoscill}. 
Here oscillatory hormone secretion from physically separated islets can be synchronized by the common stimulus of glucose.
The coordination of hormone secretion from multiple islets originates from the phase modulation responding to the common environment of glucose concentration~\cite{synch_horm, synch1,gluins}.
The entrainment through the interaction between systems and environment is an important mechanism for biological systems.
Cells or organs secrete hormones with different patterns depending on environment, and then these messengers of hormones regulate the physiological environment~\cite{hormone}.
The synchronization of Gonadotropin-releasing hormone (GnRH) neurons in the hypothalamus is another example that the GnRH pulses secreted by multiple GnRH neurons act as a common feedback stimulator~\cite{khadra2006}.

\begin{figure}[h]
\includegraphics[width=0.82\linewidth]{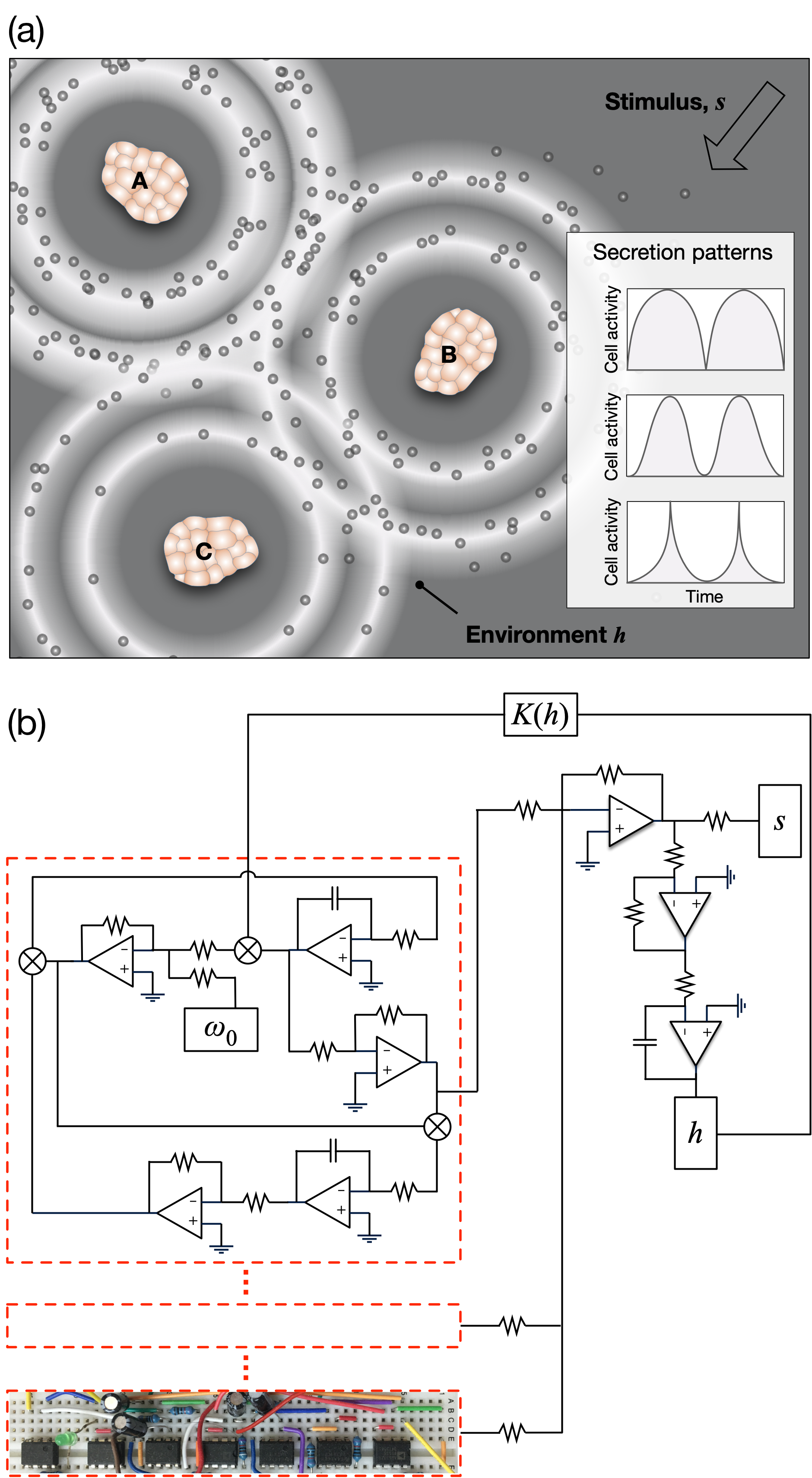}
\caption{\textbf{Biological homeostasis and electric circuit}. 
(a) The multiple elements (denoted as A, B, C) secrete messengers (small circles) responding to their surrounding environment $h$. 
The secretion patterns depend on the state of environment.
Here external stimulus $s$ and the messengers change the state of environment.
The interaction between elements and environment is controlled by a coupling strength $K(h)$.
(b) An equivalent analog electric circuit of the model system. The red dashed boxes represent each element.
In particular, the last box shows an explicit circuit with electric devices.
}
\label{fig1}
\end{figure}


Figure~\ref{fig1}(a) summarizes this mechanism of synchronization. Environment stimulates multiple components in a system (A, B, and C in Fig.~\ref{fig1}(a)), and then they secrete messengers (small circles in Fig.~\ref{fig1}(a)) that regulate the state of the environment.

We consider an active rotator as a building unit of each component that generates non-sinusoidal oscillations of which phases are modulated by the state of the environment.
The active rotator is a well-known model of limit cycle oscillators in excitable systems~\cite{AR}, which has been adopted to describe Josephson junction array, chemical reactions, charge density waves, and neuronal firing~\cite{AR_jj, AR_cr, AR_cdw, AR_nf}. 
In electric engineering, the active rotator model is also known as Adler's equation~\cite{Adler} approximating
second-order LC oscillators, and it is widely used for describing injection-locked oscillators~\cite{ilo}. 
Recently, the active rotator model is also adopted to describe the phase modulation of biological hormone secretion~\cite{ho-ar}.

The synchronization of interacting oscillators has been intensively studied, particularly for globally coupled oscillators through mean field approaches~\cite{syn1, syn2, syn3}.
In the absence of the direct coupling between oscillators, even common noise can induce synchronization between uncoupled oscillators~\cite{syn4, syn5}.
Similarly, a dynamic common environment can also induce synchronization between uncoupled periodic oscillators~\cite{syn6} and between uncoupled chaotic oscillators~\cite{syn7}. 
In this study, we propose a minimal model for the synchronization induced by the common dynamic environment. 
Unlike previous studies considering general dynamical systems~\cite{syn7} or explicitly considering amplitude and phase dynamics of oscillators~\cite{syn6}, we focus on the phase of oscillations for active rotators. 
Their phases are modulated by the state of environment, and the environment is regulated by the phases of oscillators.
Then we demonstrate this synchronization mechanism by realizing an analog electric circuit with microelectronic components, such as UA741, as shown in Fig.~\ref{fig1}(b). 
The realization of the mechanism can suggest a bio-mimetic device for coordinating multiple components to regulate environmental states.

This paper is organized as follows. 
In Sec.~\ref{sec2}, we introduce our model system, and then present the environment-dependent synchronization with the boundary of parameter space for synchrony using Ott-Antonesen ansatz~\cite{OAansatz}, which is our primary finding. 
In Sec.~\ref{sec3}, we experimentally demonstrate the synchronization mechanism. 
Here we design an analog electric circuit to realize active rotators. 
Finally, in Sec.~\ref{sec4}, we summarize our results and discuss their potential applications.

\section{Active rotators interacting with environment}
\label{sec2}
We consider a system with multiple rotators of which phases are perturbed by an environment.
The phase of the $n$-th rotator $\theta_n$ and the environment $h$ evolve with time $t$ as follows:
\begin{eqnarray}
\frac{d \theta_n}{dt} &=& \omega_n - K(h)\cos\theta_n, \label{thdot} \\
\frac{dh}{dt} &=& F( \{\theta_n \}, h, s), \label{hdot}
\end{eqnarray}
where $\omega_n$ is an intrinsic angular velocity of the $n$-th rotator, 
and $K(h)$ represents the interaction between phase $\theta_n$ and environment $h$.
The interaction strength controls the degree of phase modulation.
The first equation represents the response of rotators to environment, while the second equation describes the regulation of environment by rotators.

The regulation rate $F( \{\theta_n \}, h, s)$ could be generally dependent on the phases $\{ \theta_n \} \equiv (\theta_1, \theta_2, \dots, \theta_N)$ of every rotator, the present status $h$ of environment, and the external stimulus $s$.
As a simple but reasonable choice, we consider $F =a [s - \sum_n (1+ \cos \theta_n)]$ where the external stimulus $s$ increases the environmental variable $h$, whereas the active phases $\theta_n = (-\pi, \pi)$ of rotators decrease $h$.
The regulation term of $(1+\cos\theta_n)$ corresponds to the instantaneous area under the curve (AUC) for the phase oscillator $r_n \exp(i\theta_n)$ with a fixed amplitude ($r_n=1$). The instantaneous AUC includes a shift with the value of 1 to represent negligible (instead of negative) regulation at silent phases of $\theta_n = \pm \pi$.
Since the scale of stimulus $s$ and the amplitude of regulation rate $F$ are arbitrary, we set
\begin{equation}
F( \{\theta_n \}, h, s)= s- \frac{1}{N} \sum_{n=1}^N \cos \theta_n,
\label{hdotp}
\end{equation}
with reparameterized $s=s-1$ and $F=F/a$.
Note that the phase rotators cannot bound the increase of $h$ under too large external stimulus $s$.

We numerically solve the coupled differential equations of Eqs.~(\ref{thdot}) and (\ref{hdot}) using the fourth order Runge-Kutta method~\cite{RK4} with a sufficiently small time step, $\Delta t=0.001$.
We then demonstrate that the system-environment interaction can entrain non-interacting rotators to be synchronized.
This feedback-induced entrainment is markedly different from the unidirectional entrainment by an external oscillatory driving with a characteristic frequency $\omega$ that is not interacting with systems: $d \theta_n /dt = \omega_n + K \sin(\omega t - \theta_n)$.

\subsection{Environment-dependent synchronization}
\begin{figure}[t]
\includegraphics[width=0.82\linewidth]{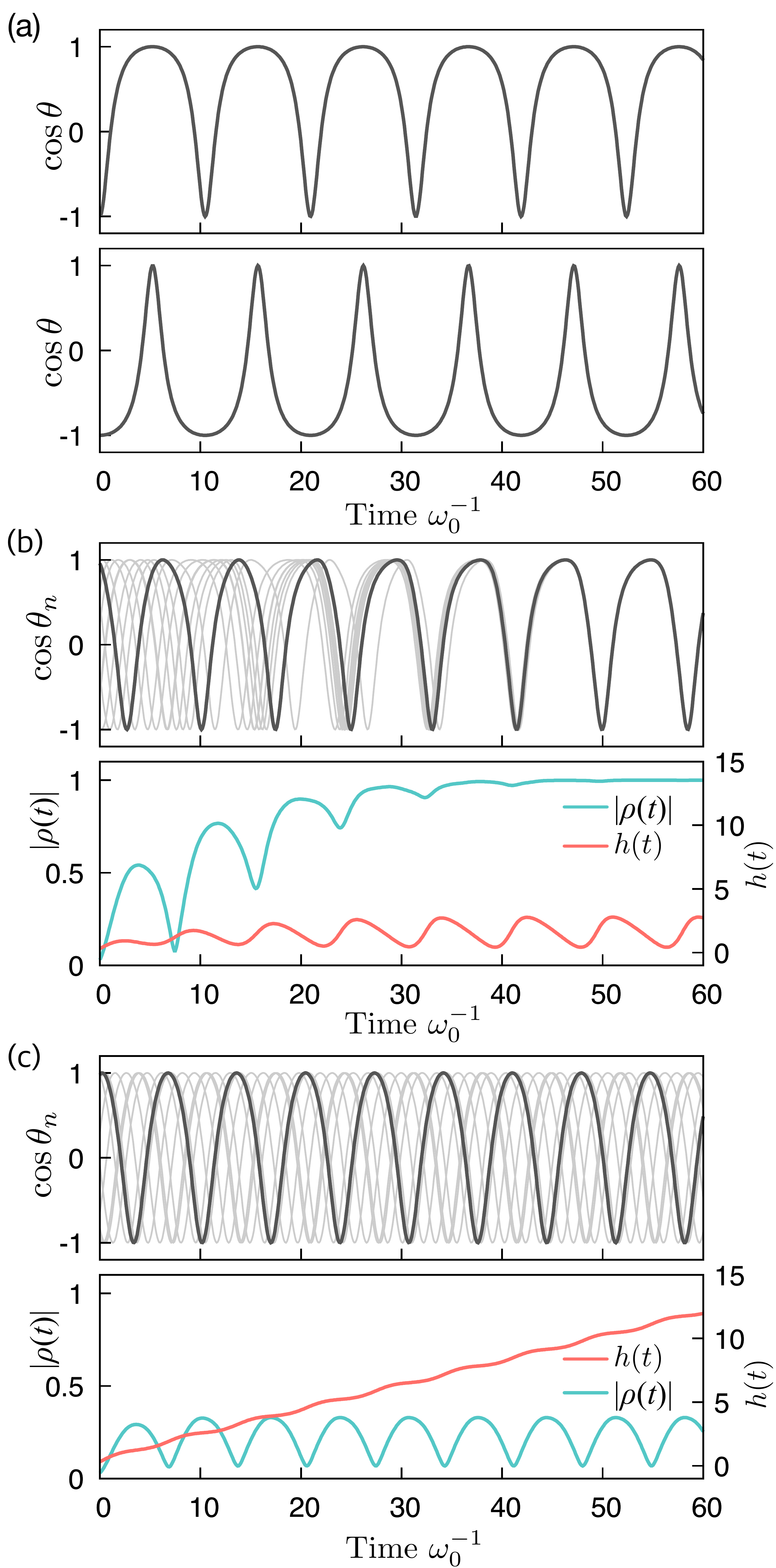}
\caption{\textbf{Environment-dependent synchronization of active rotators}. 
(a) Phase modulation of rotators depending on the modulation factor $K(h)=0.8$ (upper) and $K(h)=-0.8$ (lower).
Phase dynamics of randomly selected $20$ rotators (gray lines and one black line), their degree of synchronization ($|\rho(t)|$, green line), and the state of environment ($h(t)$, green line) for (b) a synchronizing condition ($s=0.4$, $K_0=0.8$) and (c) a non-synchronizing condition ($s=0.4$, $K_0=0.4$).
For the plot, we used $N=200$ identical oscillators ($\omega_n=\omega_0=1$) with a phase modulation function, $K(h)=K_0 \tanh h$.
}
\label{fig2}
\end{figure}

The interaction between rotators and environment is mediated by the phase modulation function $K(h)$, which is a monotonic and smoothly saturating function of $h$, e.g., $K(h)=K_0 \tanh h$.
Depending on the strength of the phase modulation, the active rotator has two regimes of distinct dynamic behavior: phase-locked and oscillatory regimes~\cite{AR}. 
Since we are interested in biological oscillation, we consider the oscillatory regime guaranteed by a constraining condition of $|K(h)| \leq K_0 \leq \omega_n$.
Given this condition, the sign of $K(h)$ determines the oscillation pattern and the ratio of active to silent phases. 
For example, given constant $K(h)=K_0$ with quenched environment ($dh/dt=0$), active rotators showed distinct oscillation patterns depending on $K_0$ (Fig.~\ref{fig2}(a)). The positive and negative plateau indicates active and silent phases, respectively.  

Once we turned on the dynamics of environment $h$, active rotators showed either complete synchronization or desynchronization depending on the interaction parameter $K_0$ and stimulus $s$. 
We numerically examined the two distinct regimes for rotators' synchrony.
Given $N = 200$ identical rotators ($\omega_n=\omega_0=1$), we explored the synchronization boundary for $K(h)=K_0 \tanh h$ and $F(\{\theta_n\},h,s)$ in Eq.~(\ref{hdotp}) by controlling the parameters $K_0$ and $s$. 
Figure~\ref{fig2}(b) shows the phase traces of randomly selected 20 rotators from total 200 rotators. 
Initial states of rotators are all different. However, as rotators interact with the common environment, they are modulated to be synchronized.
Here to probe the degree of synchronization, we used the absolute value of the complex Kuramoto order parameter, $\rho(t) \equiv \frac{1}{N}\sum_{n=1}^N\exp(i\theta_n)$~\cite{AR_cr}.
$| \rho(t) |$ initially fluctuates, continuously increases, and finally saturates at the unity representing complete synchronization.

\subsection{Synchronization boundary} 
\begin{figure}[]
\includegraphics[width=\linewidth]{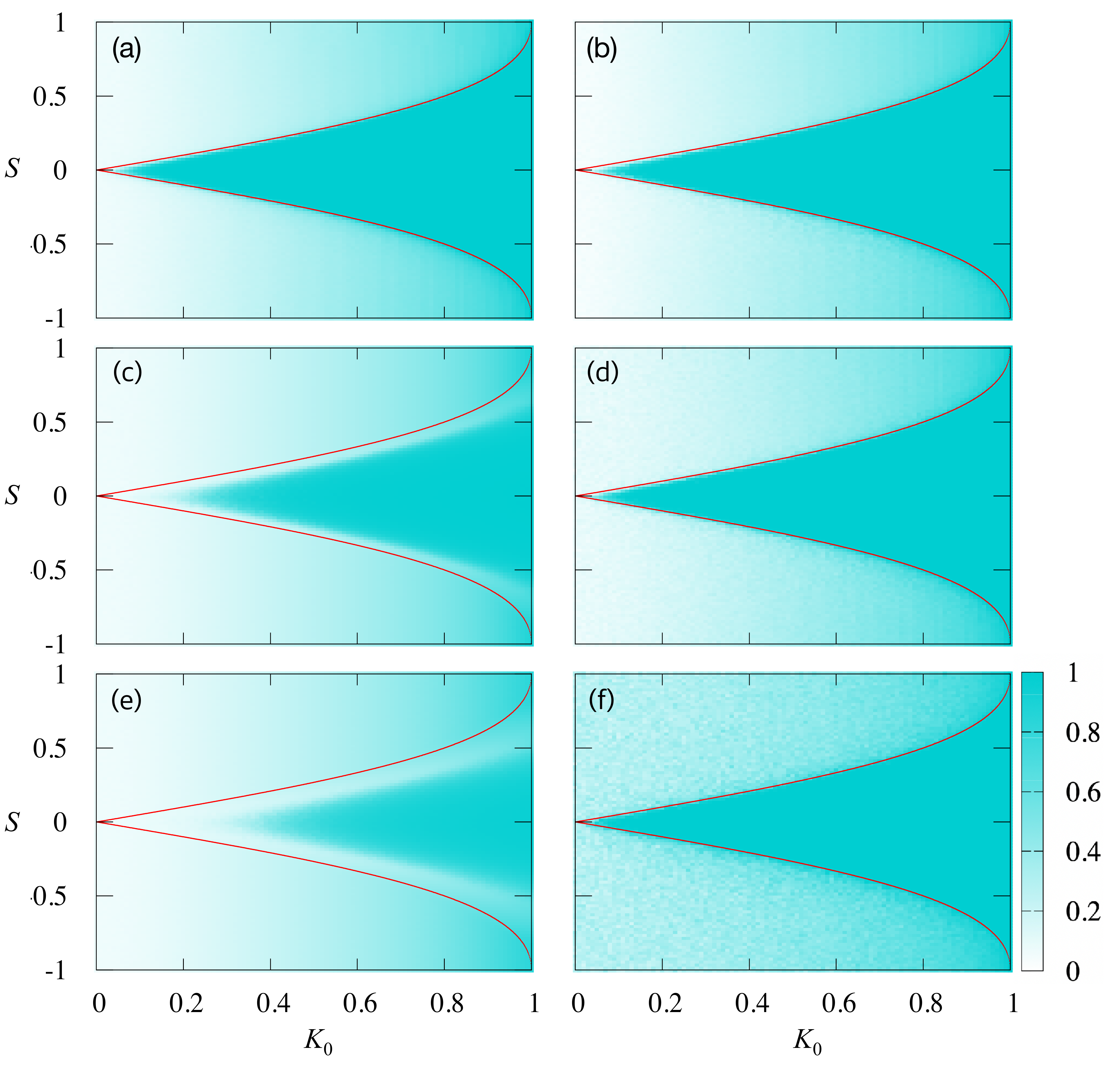}
\caption{\textbf{Boundary for complete synchronization}. 
Heat maps of degree of synchronization $|\bar{\rho}|$ for the maximum coupling strength $K_0$ and stimulus $s$.
Red line represents the theoretical synchronization boundary in Eq.~(\ref{gam}).
The boundary is robust for varying total number $N$ of oscillators and heterogeneity of their intrinsic frequencies $\omega_n$.
We sampled $\omega_n$ from a normal distribution with a mean $\omega_0=1$ and and standard deviation $\Delta \omega$.
The plots are obtained from averages of 100 ensembles for (a) ($N, \Delta \omega$)=(200, 0), (b) (500, 0), (c) (200, 0.05), (d) (100, 0), (e) (200, 0.1), and (f) (10, 0).
We numerically computed $|\bar{\rho}| \equiv\frac{\omega_0}{2\pi} \int_{T-\tau}^T |\rho(t)| dt$ with a burning period $T=1000$ and a sufficiently long period $\tau =20\pi/\omega_0$. 
}
\label{fig3}
\end{figure}

Unless the absolute level of the stimulus $|s|$ is too large, rotators always become synchronized through the dynamic feedback between rotators and environment.
In other words, if the phase modulation of rotators can manage to regulate the stimulus, the rotators are synchronized.
However, if the stimulus is too large beyond the manageable capacity of the phase rotators, the environmental variable blows up, and the rotators have drifting phases without synchronization (Fig.~\ref{fig2}(c)).
Here we obtained the threshold external stimulus $s_b$ determining the boundary for complete synchronization by using a linear stability analysis based on the Ott-Antonesen ansatz~\cite{OAansatz}:
\begin{equation}
s_{b} = \frac{\omega_0}{K_0} -\sqrt{\left(\frac{\omega_0}{K_0} \right)^2-1}, \label{gam}
\end{equation}
of which detailed derivation is referred to Appendix \ref{sec:aa}.
The synchronized area of numerical results are denoted by green area in Fig.~\ref{fig3} and the theoretical boundary for synchronization is denoted by red solid lines.
Note that the time trajectory of $|\rho(t)|$ depends on the specific shape of $K(h)$, whereas the synchronization boundary does not depend on the shape, but it depends on the saturation value $K_0$.
As shown in Fig.\ref {fig3}, we confirmed that the synchronization boundary did not change in the presence of small heterogeneity of intrinsic frequencies $\omega_n$ and under different numbers of rotators. 

\section{Experimental realization}\label{sec3}

Now we build an analog electric circuit to realize the theoretical model as shown in Fig.~\ref{fig1}(b).
The circuit mapping is straightforward by introducing new variables:
$V_{xn} \equiv\cos\theta_n$ and $V_{yn} \equiv\sin\theta_n$.
Then, the dynamics of $\dot{V}_{xn}$ can be obtained by multiplying $-V_{yn}$ to Eq.\eqref{thdot},
and $\dot{V}_{yn}$ can be similarly obtained by multiplying $V_{xn}$ to Eq.\eqref{thdot}.
Since the functional shape of $K(h)$ does not affect stationary responses of rotators, we choose a simple modulation function for experimental convenience, $K(h)=K_0 h^3$ for $h =[-1, 1]$, $K(h)=-K_0$ for $h<-1$, and $K(h)=K_0$ for $h>1$.
After changing variables with a fixed frequency $\omega_0$, Equations~(\ref{thdot}) and (\ref{hdot}) can be rewritten as follows: 
\begin{eqnarray}
\dot{V}_{xn}&=&- \big[ \omega_0-K(V_h) V_{xn} \big] V_{yn}, \label{vxdot} \\
\dot{V}_{yn}&=& \big[ \omega_0-K(V_h) V_{xn} \big] V_{xn}, \label{vydot} \\
\dot{V}_{h}&=&V_s - s_N \sum_{n=1}^N  V_{xn}, \label{vhdot}
\end{eqnarray}
where $V_h$ and $V_s$ correspond to the variables of $h$ and $s$, respectively, and $s_N$ is introduced for proper normalization. 
Then we could successfully implement an electric circuit for the theoretical model. 



\begin{figure}
\includegraphics[width=0.45\textwidth]{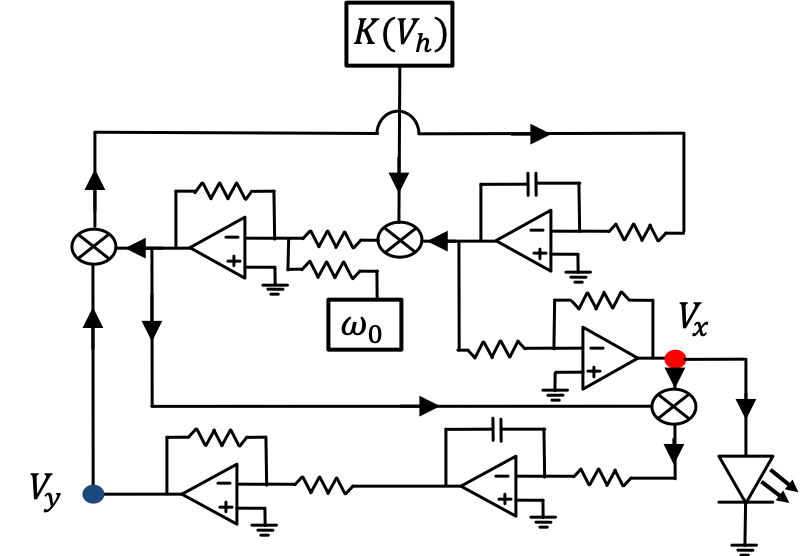}
\caption{{\bf Schematic diagram of the electric circuit for single active rotators.}
The $\omega_{0}$ and feedback signal $K(V_{h})$ from environment are considered as input parameters. 
The $\otimes$ denotes multiplier implemented by analog multiplier AD633. 
The red and blue dots indicate reference nodes for $V_{x}$ and $V_{y}$, respectively. 
}
\label{sfig1}
\end{figure}

Equations~(\ref{vxdot}) and~(\ref{vydot}) were realized on an electric circuit by using an operational amplifier (op-amp) and an analog multiplier (Fig.~\ref{sfig1}). 
The op-amps (UA741CP) were basic building blocks in our circuit design.
Except for the multiplication with an analog multiplier (AD633JN), other logic calculations were implemented by op-amp circuits for integrator, summing adder, voltage follower, and inverting amplifier~\cite{aoelec}. 
In particular, we operated integration by using the lossy integrator of ten millisecond RC time with shunt resister preventing charge storage of capacitors in the integrator.
We monitored the signals from the circuit by using Agilent oscilloscope (DSO-X 2012A) and function generator (Agilent 33220A).
Furthermore, we used green LEDs to visualize the activities of electric rotators.


\begin{figure}[]
\includegraphics[width=\linewidth]{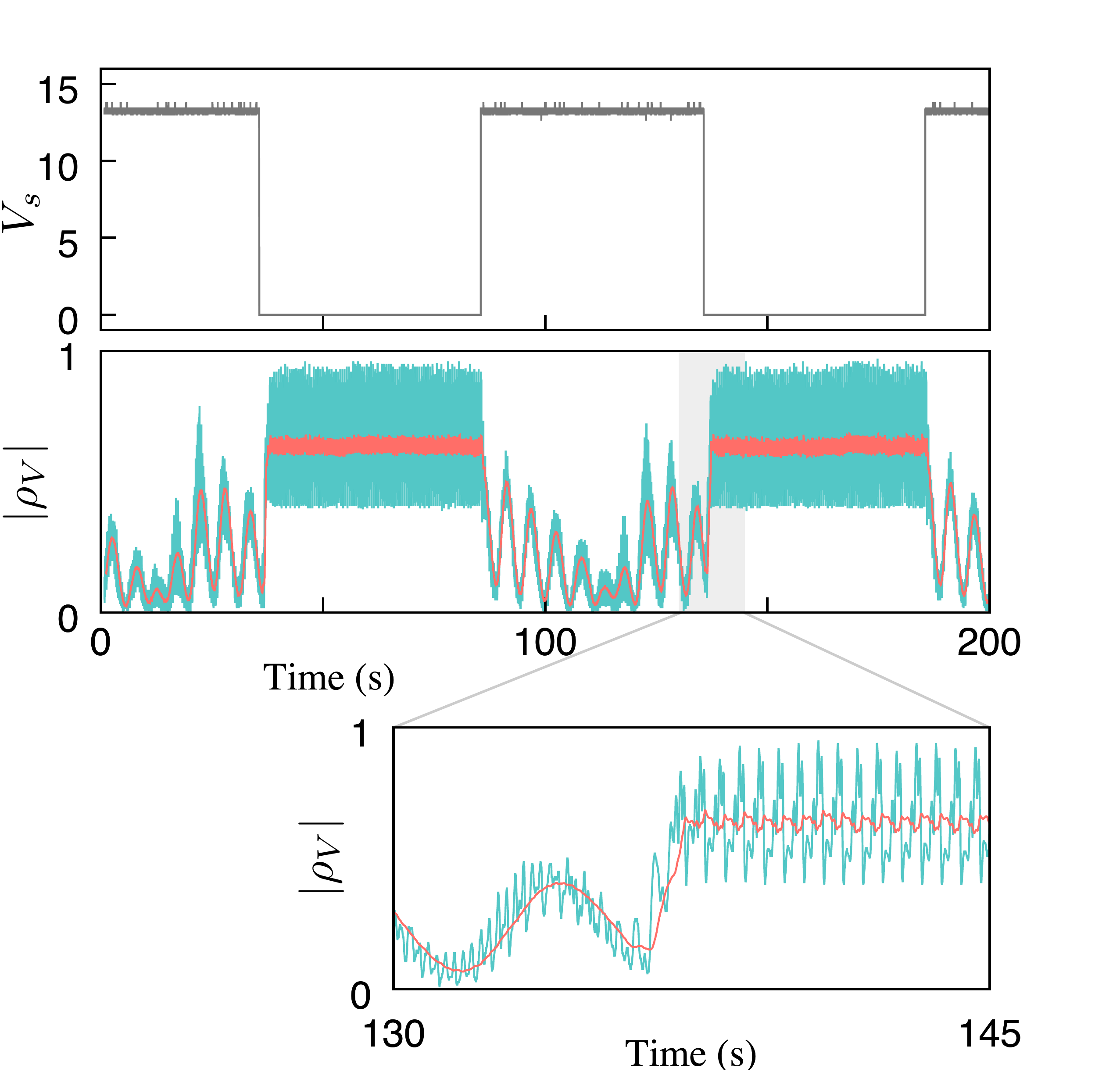}
\caption{\textbf{Controllable synchronization of electric elements}. 
Depending on the external signal $V_s$, the degree of synchronization $|\rho_V|$ changes with time (green lines).
For clear demonstration, its moving average (red line), $|\bar{\rho}_V (t)|\equiv \frac{1}{\Delta T}\int_{t-\Delta T}^{t} |\rho_V (t')| dt'$ with $\Delta T=\omega_0/2\pi\approx 0.43~\rm{s}$, is also plotted. 
The zoom-in plot of $130\leq t \leq 145$ time window shows the transition detail from desynchronization to synchronization.
}
\label{fig5}
\end{figure}

Varying system parameters such as $\omega_0$ to set a proper value of circuit components, we monitored voltage $V_{xn}$ (red filled dot in Fig.~\ref{sfig1}) and $V_{yn}$ (blue filled dot).  
Depending on the amplitude of the external stimulus $V_s$, the four electric rotators showed either synchronized or desynchronized behaviour that was measured by the order paramter $|\rho_V|$ (Fig.~\ref{fig5}).
We directly measured outputs of the trigonometric functions $V_{xn}$ and $V_{xn}$ (refer Appendix \ref{sec:ac} for the recording set-up), and computed the order parameter $|\rho_V|=\sqrt{(\sum_{n=1}^N V_{xn})^2+(\sum_{n=1}^N V_{yn})^2}/N$. 
Since $|\rho_V(t)|$ largely fluctuates for a small number $N=4$ of rotators, we used a moving average. 
For this particular demonstration, we used a fixed $K_0=0.75\omega_0$ and two values of stimulus ($s=0$ for synchronization and $s=15\omega_0$ for desynchronization), and set the natural frequency $\omega_0 / 2\pi$ corresponding to $2.33~\rm{Hz}$.

\section{Discussion} \label{sec4}
 
Synchronization of oscillators has been extensively studied in various contexts including biological~\cite{bio_synch} and engineering systems~\cite{grid_synch}. 
The control of synchronization has been mainly achieved by changing the coupling strength between oscillators~\cite{AR}. 
In this study, however, we considered interactions between systems and environment as a natural way to induce synchronization between non-interacting elements in a system.
The active system-environment feedback has been proved to be useful for the adaptation of robot locomotion~\cite{AR_robot}.
To control the robot locomotion, Owaki and colleagues have considered the interaction between robot legs and local reaction force from ground (environment).
The motion of legs has been modeled by active rotators: $d \theta_n /dt = \omega - K(h_n) \cos \theta_n$, 
while the local reaction force $h_n$ for each leg should depend on the posture of the four legs with different phases such as $h_n (\theta_1, \theta_2, \theta_3, \theta_4)$.
Unlike the heterogeneous local environment $h_n$, our model considered a homogeneous global enviromnent $h$.

Bio-mimetic devices have been emphasized with their advanced functions of redundancy, low power, high sensitivity, and multiple purposes~\cite{biomimetics}.
PID controller~\cite{PID} is a state-of-the-art technology as a closed loop controller to maintain a desirable set point. 
Inspired by the biological homeostasis, the biological mechanism may propose a bio-mimetic device for controlling set point.
Unlike the single-unit PID controller, our model suggested that the phase coordination of multiple units could be another mechanism for regulating environment in addition to the amplitude modulation of single units.
The synchronization response of multiple units could be used as a sensor for monitoring varying environment, and also as an amplification of signals for regulating environment.



In summary, we proposed a simple model for describing phase coordination between multiple rotators influenced by environment.
Based on the closed loop interaction between the environment and multiple rotators, we found that the dynamic environment could entrain non-interacting rotators if the phase responses of rotators could manage external perturbation on environment.
We analyzed the synchronization boundary depending on the environment-system coupling strength $K_0$ and the level of external perturbation $s$, and showed that either synchronization or desynchronization regime existed with clear separation through the boundary.
Moreover, we realized the synchronization mechanism using an electric analog circuit.
The circuit can be potentially applicable for practical purposes as an analog controller, and it can serve as a bio-mimetic platform to further understand the regulation of biological oscillation.

\begin{acknowledgments}
This research was supported by the Basic Science Research Program through the National Research Foundation of Korea (NRF), funded by the Korea government (MSIT) through NRF-2019R1F1A1052916 (J.J.), and by the Ministry of Education through NRF-2017R1D1A1B03032864 (S.-W.S.), and by the Ministry of Science, ICT $\&$ Future Planning through NRF-2017R1D1A1B03034600 (T.S.). 
\end{acknowledgments}

\setcounter{figure}{0}
\renewcommand{\thefigure}{A\arabic{figure}}
\appendix
\section{Linear stability analysis}  \label{sec:aa}

We derive the dynamic equation to describe the degree of synchronization between the active rotators.
The phase dynamics of the active rotators is
\begin{equation}
\frac{d \theta_n}{dt} = \omega_n - K \cos \theta_n \
\label{eq:ar}
\end{equation}
for $n = 1, \cdots, N$.
In the continuum limit of $N \to \infty$, we consider the instantaneous phase distribution $P(\omega, \theta, t)$ with the normalization condition $\int_0^{2 \pi} d\theta P(\omega, \theta, t) = 1$.
The probability density satisfies the Fokker-Planck equation,
\begin{equation}
\label{eq:FP}
\frac{\partial P}{\partial t} = - \frac{\partial}{\partial \theta} \bigg[ \bigg( \omega - K \frac{e^{i \theta}+e^{-i \theta}}{2} \bigg) P \bigg].
\end{equation}
Using $P(\omega, \theta, t)$, we can define the order parameter $\rho(t)$ that measures the degree of synchronization between rotators,
\begin{equation}
\label{eq:rho}
\rho(t) \equiv \int_{-\infty}^{\infty} d\omega \int_0^{2\pi} d\theta P(\omega, \theta, t) e^{i \theta}.
\end{equation}
To obtain the dynamics of $\rho(t)$, we use the Ott-Antonsen ansatz~\cite{OAansatz},
\begin{equation}
P(\omega, \theta, t) = \frac{g(\omega)}{2 \pi} \bigg[ 1 + \sum_{m=1}^\infty \bigg( \alpha^m(\omega,t) e^{i m \theta} + \bar{\alpha}^m(\omega,t) e^{-i m \theta} \bigg) \bigg],
\end{equation}
where $g(\omega)$ is the distribution of intrinsic frequency $\omega$, and $\bar{\alpha}(\omega, t)$ is the complex conjugate of $\alpha(\omega, t)$.
Putting this ansatz into the above Fokker-Planck equation in Eq.~(\ref{eq:FP}), we obtain the following two equations:
\begin{align}
\frac{\partial \alpha}{\partial t} &= - i \omega \alpha + i \frac{K}{2} (1 + \alpha^2), \\
\label{eq:bar}
\frac{\partial \bar{\alpha}}{\partial t} &=  i \omega \bar{\alpha} - i \frac{K}{2} (1 + \bar{\alpha}^2),
\end{align}
by considering the independent $m$-th order for $e^{im\theta}$ and $e^{-im\theta}$. 

It is straightforward to show $\rho(t) = \bar{\alpha}(t)$ from Eq.~(\ref{eq:rho}), given the identical intrinsic frequency of $g(\omega) = \delta(\omega-\omega_0)$.
Therefore, $\rho(t)$ should be also governed by Eq.~(\ref{eq:bar}) as
\begin{equation}
\frac{d \rho}{d t} =  i \omega_0 \rho - i \frac{K}{2} (1 + \rho^2).
\end{equation}
Note that the partial derivative for time $t$ is changed to the total derivative because now $\rho$ is dependent only on $t$.
This equation for the complex variable $\rho(t) = x(t) + i y(t)$ can be decomposed into the equations for real variables, $x$ and $y$:
\begin{align}
\frac{dx}{dt} &= -(\omega_0 - K x) y, \\
\frac{dy}{dt} &= \omega_0 x - \frac{K}{2} (1 + x^2 - y^2).
\end{align}
Indeed, we are interested in the amplitude $|\rho(t)|$ of the order parameter.
Its squared value $|\rho|^2 = x^2 + y^2$ evolves as
\begin{equation}
\frac{d|\rho|^2}{dt} = 2x \frac{dx}{dt} + 2y \frac{dy}{dt} = -K y (1-|\rho|^2).
\end{equation}
Since the coupling parameter $K$ is defined as a positive value, the stationary solution for this equation is $y=0$ or $|\rho|=1$.
Later we will show that the solution $y=0$ cannot be stable.
However, $\rho(t) \le 1$ largely fluctuates for $|\rho| \ll 1$, whereas it minimally fluctuates when $|\rho| \approx 1$.
Thus, we expect that $\rho(t)$ evolves to the stationary solution $|\rho|=1$ that represents the complete synchronization between rotators.

Now we consider that the coupling parameter $K=K(h)$ depends on the environmental status $h$.
The variable $h$ is governed by an external source $s$ and the feedback from active rotators:
\begin{equation}
\frac{dh}{dt} = s - \frac{1}{N} \sum_{n=1}^N \cos \theta_n.
\end{equation}
Here the feedback term is the real part $x \equiv  \frac{1}{N} \sum_{n=1}^N \cos \theta_n$ of $\rho \equiv \frac{1}{N} \sum_{n=1}^{N} e^{i \theta_n}$.
Then, we obtain complete equations for $(x, y, h)$ as follows
\begin{align}
\label{eq:x}
\frac{dx}{dt} &= - \bigg( \omega_0 - K(h) x \bigg) y, \\
\label{eq:y}
\frac{dy}{dt} &= \omega_0 x - \frac{K(h)}{2} (1 + x^2 - y^2),\\
\label{eq:h}
\frac{dh}{dt} &= s -x.
\end{align}
The above stationary solution $|\rho(t)| = 1$ assumed that $K$ is constant.
We examine if $K(h)$ can be constant with $dh/dt=0$ for the given solution $|\rho(t)| = 1$.
Under the full synchronization $(\theta_n = \theta)$, Eq.~(\ref{eq:h}) becomes $dh/dt = s - \cos \theta$.
Then, its time-averaged equation for a period $T$ of oscillations is
\begin{equation}
\bigg\langle \frac{dh}{dt} \bigg\rangle = s - \langle \cos \theta \rangle,
\end{equation}
for a constant stimulus $s$. Here the time average is defined as $\langle f \rangle \equiv \frac{1}{T} \int_0^T f(t) dt$.
The identical active rotators have a time period,
\begin{align}
T &= \int_0^{2\pi} \frac{d\theta}{d\theta/dt} = \int_0^{2\pi} \frac{d\theta}{\omega_0 - K \cos \theta} \nonumber \\
&= \frac{2\pi}{\sqrt{\omega_0^2 - K^2}} .
\end{align}
Given $T$, the time average of $\cos \theta(t)$ is
\begin{align}
\langle \cos \theta \rangle &= \frac{1}{T} \int_0^T \cos \theta(t)~dt \nonumber \\
&= \frac{\omega_0}{K} - \sqrt{ \bigg( \frac{\omega_0}{K} \bigg)^2 -1}.
\end{align}
Then, the manageable positive stimuli $s$ by the averaged response from the rotators should satisfy the following inequality,
\begin{equation}
s \le \langle \cos \theta \rangle \le \frac{\omega_0}{K_0} - \sqrt{ \bigg( \frac{\omega_0}{K_0} \bigg)^2 -1},
\end{equation}
which guarantees the stationarity of $\langle dh/dt \rangle =0$.
This condition explains the synchronization boundary in Fig. 3 in the main text.

Finally, we explore the other possible stationary solutions for $(x, y, h)$ of Eqs.~(\ref{eq:x})-(\ref{eq:h}).
Suppose that other stationary solutions $(x^*, y^*, h^*)$ exist with the stationarity conditions ($dx/dt = dy/dt = dh/dt = 0$).
(i) The first condition $dx/dt=0$ implies $x^* = \omega_0/K$ or $y^*=0$, where $x^*=\omega_0/K >1$ cannot be a solution due to $|x| \le |\rho| \le 1$, given $\omega_0 > K$.
(ii) The second condition $dy/dt=0$ with $y^*=0$ implies 
\begin{equation}
\label{eq:xh}
x^* = \frac{\omega_0}{K(h^*)} - \sqrt{\bigg( \frac{\omega_0}{K(h^*)} \bigg)^2-1},
\end{equation}
where another solution $x^* = \omega_0/K(h^*) + \sqrt{(\omega_0/K(h^*))^2-1}$ is excluded due to the condition $|x| \le 1$.
(iii) The third condition $dh/dt=0$ finally imposes $x^*=s$ that fixes $x^*$ and $h^*$ in Eq.~(\ref{eq:xh}).
Is this solution $(x^*, y^*, h^*)$  stable?
To examine its stability, we consider $x = x^* + \epsilon_x$, $y = y^* + \epsilon_y$, and $h = h^* + \epsilon_h$, located slightly away from the fixed point $(x^*, y^*, h^*)$.
Then, the time evolution of the deviation vector ${\bf{\epsilon}}=(\epsilon_x, \epsilon_y, \epsilon_h)$ can be derived up to their linear orders as $d{\bf{\epsilon}}/dt = {\bf J \epsilon}$ by using Eqs.~(\ref{eq:x})-(\ref{eq:h}).
The Jacobian matrix is defined as
\[
   {\bf J}=
  \left[ {\begin{array}{ccc}
   0 & -\omega_0 + K(h^*)x^* & 0 \\
   \omega_0 - K(h^*)x^* & 0 & \frac{1}{2}\big[ \frac{dK}{dh} \big]^* (1+x^{*2}) \\
   -1 & 0 & 0\\
  \end{array} } \right]
\]
with the derivative $[dK/dh]^*$ at $h=h^*$.
The eigenvalues of ${\bf J}$ can be obtained from the equation of $|{\bf J} - \lambda {\bf I}|=0$:
\begin{eqnarray}
& \lambda^3 &+ \big( \omega_0 - K(h^*)x^* \big) ^2 \lambda   \nonumber \\
 &+& \big(\omega_0 - K(h^*)x^* \big)\frac{1}{2}\bigg[ \frac{dK}{dh} \bigg]^* (1+x^{*2}) = 0.
\label{eq:lambda}
\end{eqnarray}
This polynomial equation of $\lambda^3 + a \lambda + b = 0$ for $a, b > 0$ should have one real negative eigenvalue ($\lambda_1 < 0$) and two complex eigenvalues ($\lambda_{2, 3} = \alpha \pm i\beta$), where the real value $\alpha = -\lambda_1/2 > 0$ must be positive.
Therefore, the deviation of ${\bf \epsilon}$ cannot be vanished with time. 
In other words, the stationary solution ($x^*, y^*, h^*$) cannot be stable.
However, once $[dK/dh]^*=0$, Eq.~(\ref{eq:lambda}) becomes $\lambda^3 + \big(\omega_0 - K(h^*)x^* \big)^2 \lambda =0$.
Then, the eigenvalues of ${\bf J}$ are $\lambda_1=0$, $\lambda_{2, 3} = \pm i\sqrt{\omega_0^2 - K^2(h^*)}$.
Indeed, the saturation condition $[dK/dh]^*=0$ represents $K(h^*) = K_0$.
This implies the existence of an oscillatory solution around $(x^*, y^*=0, h^*)$ with an effective frequency $\omega_{\text{eff}} \equiv \sqrt{\omega_0^2 - K^2_0}$.
The third stationary condition $x^* = s$ with Eq.~(\ref{eq:xh}) reproduces the synchronization boundary, $s={\omega_0}/{K_0} - \sqrt{ ( {\omega_0}/{K_0} )^2 -1}$, again.

\section{Experimental set-up} \label{sec:ac}

\begin{figure}[h]
\includegraphics[width=0.45\textwidth]{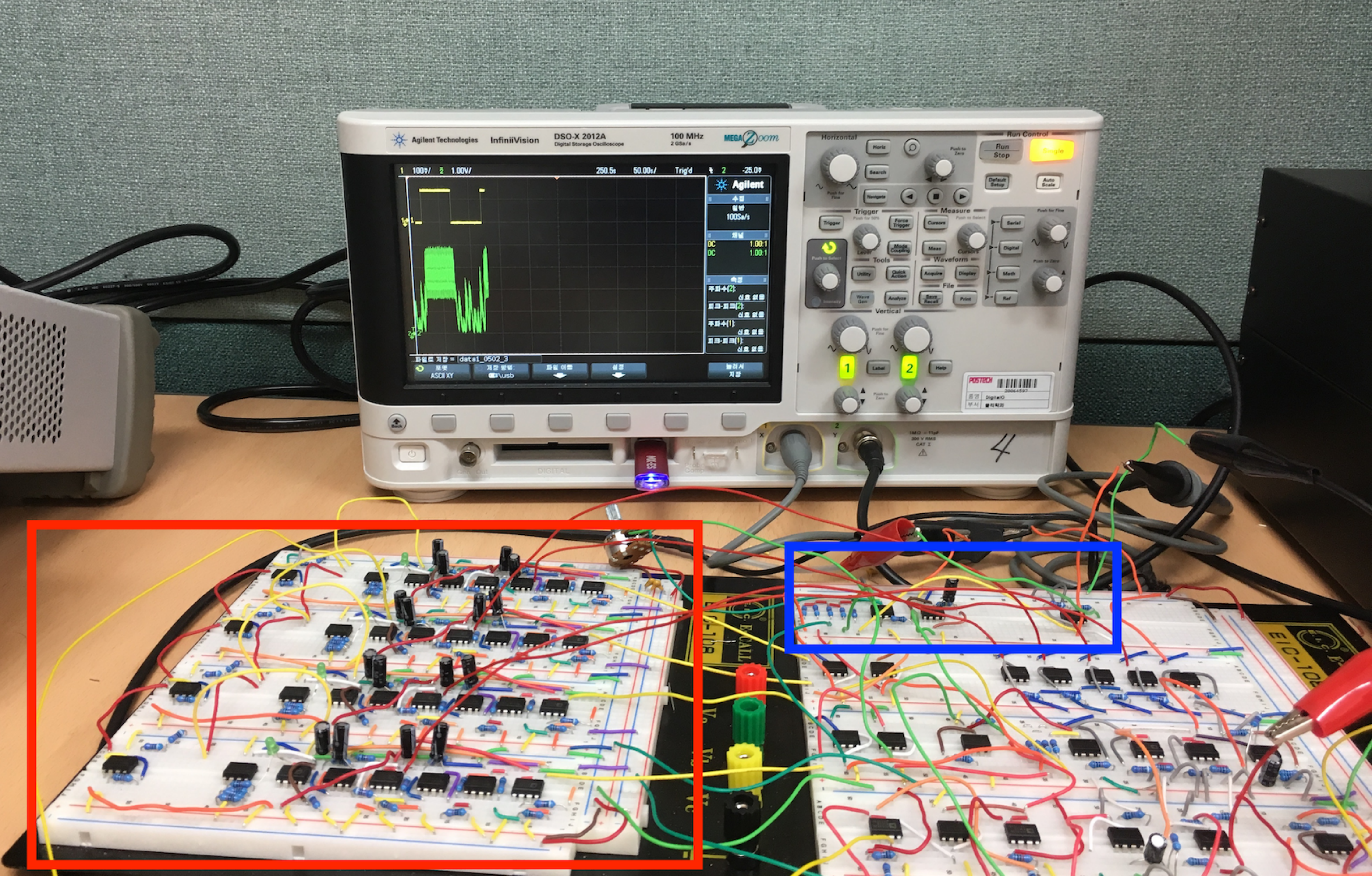}
\caption{{\bf Recording set-up}.
A power supplier provides $\pm15$ V to the breadboard during the experiment, while a function generator generates controlling voltage $V_{s}$.
The oscilloscope panel shows two signals of $V_s$ (yellow line) and $|\rho_V|$ (green line) from the breadboard.
}
\label{sfig2}
\end{figure}

The system consists of two parts:
(i) electric rotators, $V_{xn}$ and $V_{yn}$ (red boxed area in the Fig.~\ref{sfig2}) and (ii) environment, $V_h$ (blue boxed area). 
We put four copies of the electric rotators, and then connected to the environment. 
To easily monitor the degree of synchronization between four rotators, we put four green LEDs as shown in Figs.~\ref{sfig1} and \ref{sfig2}. 
The LED lights were on if voltages over $0.7$ V were applied. The threshold voltage corresponds to $1$ V for the electric rotators.

\bibliography{bibinfo_pre}

\end{document}